\documentclass[sn-mathphys]{sn-jnl}
\usepackage{array}
\usepackage[caption=false,font=normalsize,labelfont=sf,textfont=sf]{subfig}
\usepackage{textcomp}
\usepackage{stfloats}
\usepackage{url}
\usepackage{verbatim}
\usepackage{graphicx}
\usepackage{hyperref}
\usepackage{amsthm}
\usepackage{float}
\usepackage{wrapfig}

\jyear{2022}%

\theoremstyle{thmstyleone}%
\theoremstyle{thmstyletwo}%

\theoremstyle{thmstylethree}%

\begin{document}

\title[Enhance Ambiguous Community]{Enhance Ambiguous Community Structure via Multi-strategy Community Related Link Prediction Method}


\author[1,2]{\fnm{Qiming} \sur{Yang}}\email{asdyqm@buaa.edu.cn}

\author*[1,2,3,4]{\fnm{Wei} \sur{Wei}}\email{weiw@buaa.edu.cn}

\author[1,2]{\fnm{Ruizhi} \sur{Zhang}}\email{ruizhiz@buaa.edu.cn}

\author[1,2]{\fnm{Bowen} \sur{Pang}}\email{pangbw@buaa.edu.cn}

\author[5]{\fnm{Xiangnan} \sur{Feng}}\email{fengxiangnan@gmail.com}

\affil[1]{\orgdiv{School of Mathematical Sciences}, \orgname{Beihang University}, \orgaddress{\street{XueYuan Road No.37}, \city{Beijing}, \postcode{100191}, \state{Beijing}, \country{China}}}

\affil[2]{\orgdiv{Key Laboratory of Mathematics Informatics Behavioral Semantics}, \orgname{Ministry of Education}, \orgaddress{\city{Beijing}, \country{China}}}

\affil[3]{\orgdiv{Institute of Artificial Intelligence}, \orgname{Beihang University}, \orgaddress{\street{XueYuan Road No.37}, \city{Beijing}, \postcode{100191}, \state{Beijing}, \country{China}}}

\affil[4]{\orgdiv{Zhongguancun Laboratory}, \orgaddress{\city{Beijing}, \country{China}}}

\affil[5]{\orgdiv{the Center for Humans and Machines}, \orgname{Max Planck Institute for Human Development}, \orgaddress{\street{Lentzeallee 94}, \city{Berlin}, \postcode{14195}, \country{Germany}}}


\abstract{Most real-world networks suffer from incompleteness or incorrectness, which is an inherent characteristic of real-world networks. This imperfection might cause the deficiency of machine learning algorithms in complex networks like community detection. Therefore, proper measures for improving community detection performance and robustness are needed for more promising network analysis. In this paper, we propose a \underline{h}armony-based \underline{a}ggregation \underline{p}referred (HAP) link prediction method, which takes the result of community detection algorithm into consideration. The HAP method aims to detect boundary nodes among different communities and establish connections to clarify community structure. Furthermore, we design a two-step community enhancement framework with an automatic evolution process based on the HAP method. This methodology successfully clarifies ambiguous community structures by adding links to the current network. The experimental results on twelve real-world datasets with ground truth knowledge indicate that the proposed link prediction method outperforms other baseline methods in enhancement capacity and numerical stability. Furthermore, the proposed community enhancement method follows the expected evolution process.}




\maketitle
\section{Introduction}\label{sec:introduction}

Since a great deal of real-world data could be expressed in complex network fashion \citep{Strogatz(2001)networks}, complex network analysis attracts more and more attention in many scientific disciplines. There are various approaches to unveiling the underlying information behind networks. Community detection has been considered as one of the most vital among these studies \citep{deng(2016)mobility, wang(2013)discovering, qiao(2020)dynamic}. The network community is defined as a group of nodes that are densely connected to each other while sparsely connected to the rest nodes \citep{Newman(2004)community}. Community structure appears at a high frequency and is of pivotal importance in network analysis.  Social networks are paradigmatic examples of graphs with communities since people tend to form groups with similar interests or ideologies \citep{Girvan(2002)community}. In biological bodies, community structure in protein interaction networks could represent a group of proteins with similar functions \citep{Lewis(2010)function}. 

Unfortunately, most real-world datasets are severely incomplete \citep{Fortunato(2010)community}. Online social networks like Facebook and Twitter, only a tiny part of the information can be collected. In gene interaction networks, links among genes are measured by costly experiments. The imperfection of real-world network datasets consistently leads to incorrect community detection outcomes. As a fundamental network analysis method, community detection methods' skewed results will adversely impact downstream network analysis tasks \citep{Fortunato(2016)guide}. This circumstance leads to the critical need for community enhancement methods, which stands for rewiring the current network to improve the performance of existing community detection methods.

There are several challenges when designing community enhancement algorithms. As a preprocessing method, it should have the property of highly accessible computation cost. In addition, the model should not contain too many parameters since additional time is needed to learn these parameters for different graphs. Moreover, the process of enhancing community structure should be gentle and gradual, which is conducive to maintaining the algorithm's stability and genericity. From this point of view, similarity-based link prediction methods come into our sight. As network augmentation methods, they have the merit of low computation cost, guaranteeing their application on large-scale networks for multiple rounds \citep{Wang(2015)lpsurvey}.

Research has been conducted to fix impaired network systems by predicting which node pairs are more likely to establish links, also known as the link prediction method \citep{Lu(2011)linkprediction, liben(2007)CN, Cannistraci(2013)CAR, Adamic(2003)AA, Ravasz(2002)HDI, Katz(1953)KI, Li(2011)MERW}. However, to the best of our knowledge, research on discovering new link prediction methods to enhance community structure has rarely been discussed. There is adequate evidence that we could expect link prediction methods to be implemented as a data augmentation procedure in the preprocessing process of community detection methods. It could play a non-negligible role in community detection tasks as the preprocessing process has been proven a critical part of the practice of machine learning algorithms \citep{garcia(2015)data, lu(2016)preprocessing}.

\begin{figure*}[tb]
	\vspace{-1em}
	\centering  
	\includegraphics[width=\textwidth]{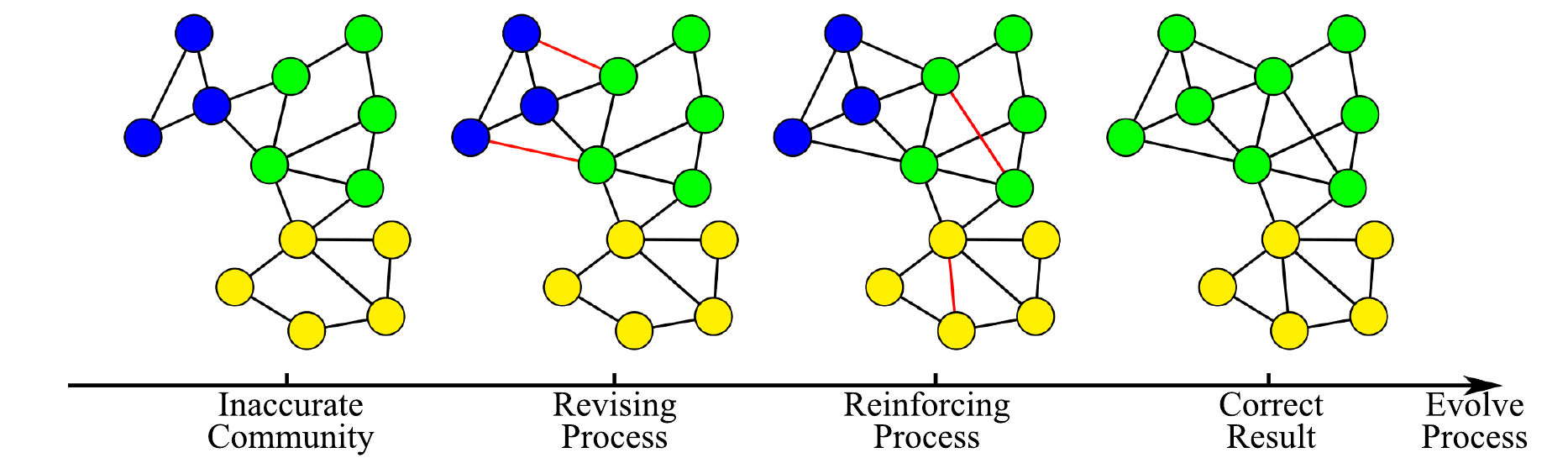}  
	\caption{Diagram of proposed community enhancement measure. The first subfigure reveals the biased community detection result by algorithms. Furthermore, the last subfigure illustrates the ground truth community. In the revising process the community enhancement measure adjusts biased result by adding inter-cluster links between inaccurately separated subcommunities. On the contrary, it enhances the correct result by adding intra-cluster edges in the reinforcing process.} 
	\label{diagram}
	\vspace{-1em}
\end{figure*}

In this paper, we design a harmony-based aggregation preferred (HAP) link prediction method and propose a community enhancement algorithm based on this strategy. It has the desirable characteristics of feasible computation cost and universal applications. Our method comprises two distinct procedures. The revising process will suture fractured communities into a complete one, while the repaired communities will get augmented during the reinforcing process. The diagram of such a procedure is presented in figure \ref{diagram}. Experimental results on real-world datasets with ground truth community information show that our method performs better than baseline methods in most cases. The main contributions of our work are threefold:
\begin{itemize}
	\item We analyze the typical bias of community detection methods due to the incompleteness of networks, and we put forward the definition of revising and reinforcing edges. Both of them will help other researchers when designing their algorithms.
	\item We design a new parameter-free, unsupervised link prediction strategy HAP for community enhancement tasks. The HAP method has low computational complexity and high universality of appliances. It could be easily applied as a plug-in module in preprocessing procedures for arbitrary community detection algorithms.
	\item We conduct experiments on 12 datasets from different areas. The extensive experimental results show that our proposed method can achieve promising performance and outperform all baseline methods.
\end{itemize}

The remainder of this paper is organized as follows. Section \ref{sec:greetings} presents related research work. Next, section \ref{sec:methods} provides the formal definition of the community detection problem and illustrates the details of our method, including the inductive biases. Furthermore, section \ref{sec:expriment} shows the experimental results and comparison with other baseline models. Finally, section \ref{sec:conclusion} gives the conclusion and discussion on future work.

\section{Related Work}
\label{sec:greetings}
Both link prediction and community detection are of great significance in network analysis since they provide network topology information from various perspectives. 

Several community detection algorithms were proposed based on altering network topology structures. Zhang et al. \citep{zhang(2013)enhanced} designed an enhanced semi-supervised learning framework for community detection, but it required prior knowledge about nodes. Yang et al. \citep{yang(2015)active} considered which prior information is critical for performance improvement and proposed an active link selection framework. Su et al. \citep{Su(2019)enhance} proposed CSE method based on central and boundary node identification for community enhancement, which successfully removed the limitation of prior knowledge about nodes. Zhou et al. \citep{zhou(2021)robustecd} proposed genetic algorithm and similarity ensemble-based community enhancement methods to explore the robustness under adversarial attack. For a big picture about community detection methods, readers are recommended to this comprehensive survey  \citep{Fortunato(2010)community}.

For future reference, three well-known and representative community detection algorithms are introduced here:
\begin{itemize}
	\item Label Propagation (LPA) \citep{Raghavan(2007)lpa}: LPA method solely uses the network structure as its information, with each node adopting the label that most of its neighbours currently have at every iteration step. The significant advantage of this algorithm is that it has near-linear time complexity.
	
	\item Infomap \citep{Rosvall(2008)infomap}: This method uses the probability flow of random walks on a network as a proxy for information flows in the real-world system. It is an information-based approach capable of revealing community structure in weighted and directed networks.
	
	\item Louvain \citep{Blondel(2008)louvain}: It is a heuristic method based on modularity optimization. This algorithm first assigns different community labels to all nodes and then optimizes the modularity by aggregating those separate communities.
\end{itemize}

In link prediction oriented problems, the community detecting results could be regarded as a global attribute to provide extra information for link prediction algorithms. Soundarajan and Hopcroft \citep{Soundarajan(2012)community} rewrote the classic CN index and RA index with community information, and the experimental results showed improvement. Rebaza and Lopes \citep{Rebaza(2012)cluster} took intra-cluster and inter-cluster into consideration and proposed WIC measure, which can be extended on directed and asymmetric large-scale networks \citep{Rebaza(2012)Twitter}. Ai et al. \citep{Ai(2019)recommend} presented a link prediction method based on complex network modelling and community detection results for personalized recommendation circumstances.

For the main focus of this paper, link prediction methods can be implemented to enhance ambiguous community structure. Yang et al. \citep{Yang(2009)link_content} proposed a conditional model for link prediction and a discriminative model for content analysis. Chen et al. \citep{Chen(2016)enhance} tested three traditional link prediction methods for enhancing community structure. Bacco et al. \citep{Bacco(2017)multilayer} proposed a generative model for multilayer networks with interdependence among their layers. Jiang et al. \citep{Jiang(2020)ambiguous} designed a strategy based on node centralities to establish clear boundaries among communities. Burgess et al. \citep{Burgess(2016)consensus} proposed EdgeBoost structure and explored the improvement of community detection performance of three link prediction algorithms with six community detection methods. 

\begin{figure*}[hb]
	\vspace{-1em}
	\centering  
	\includegraphics[width=0.9\textwidth]{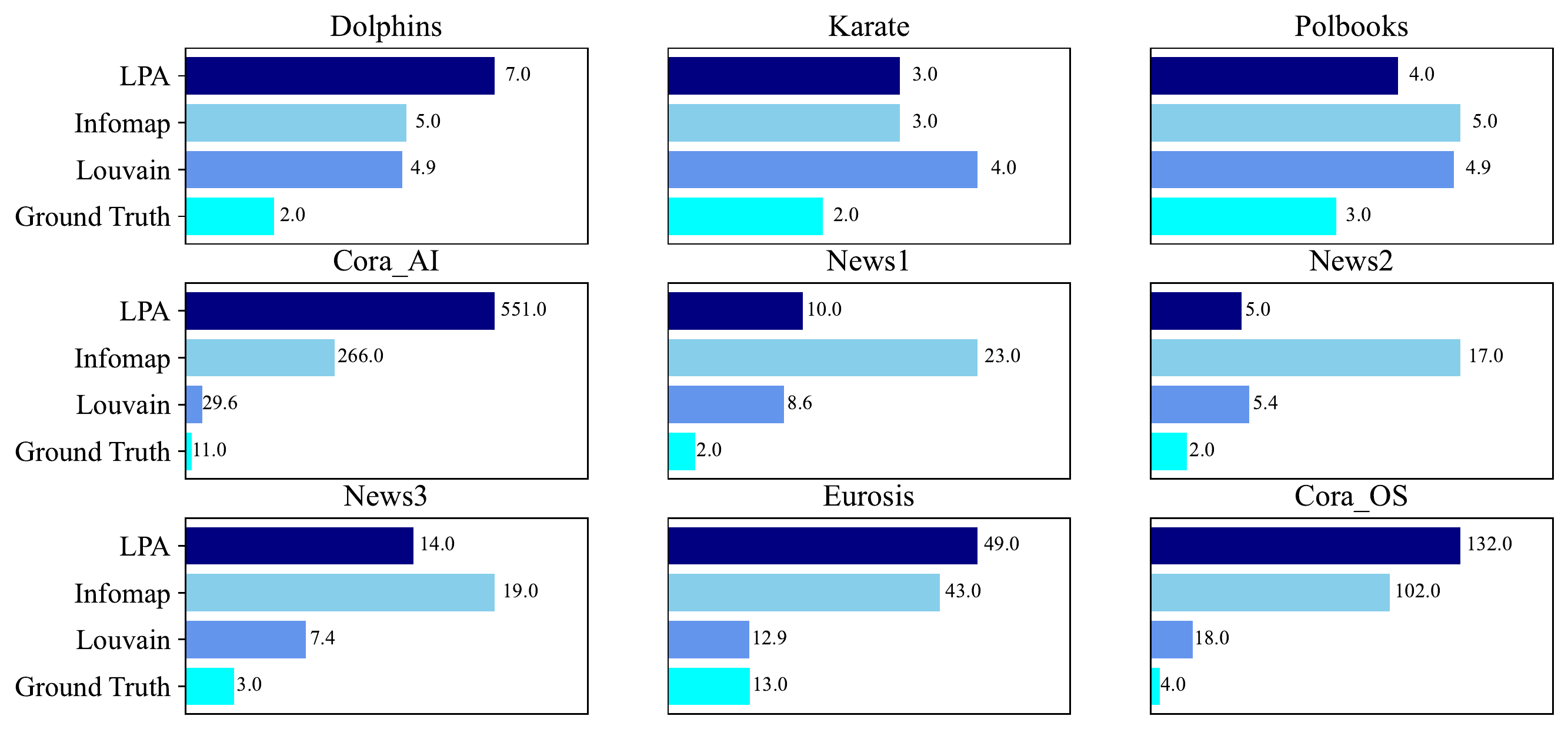}  
	\caption{The average number of clusters of community detection algorithms on nine network datasets. Each value is the average number of clusters in 10 independent iterations. Compared with the ground truth, LPA and Infomap community detection methods deliver a higher number of communities. On the other hand, the Louvain algorithm tends to detect relatively fewer communities than those two methods above but is still biased except for the Eurosis dataset.} 
	\label{nine dataset}
	\vspace{-1em}
\end{figure*}

\section{Methods}\label{sec:methods}
In this section, we explain the proposed community enhancement algorithm in detail. The critical component of the proposed enhancement algorithm is the HAP link prediction method, based on which the inductive biases and the intuitions will be illustrated. Firstly we will formally define the problem and all the symbols used in this paper.

\subsection{Problem and Definitions}
Every complex network system can be presented as an ordered tuple $G=(V, E)$, where $V$ is the set of nodes and $E\subseteq V\times V$ represents the edge set of network $G$. In this paper we mainly concern about undirected graph, i.e., $\forall v_i,v_j \in V , \ (v_i,v_j) \in E \Rightarrow (v_j,v_i) \in E$. 

Each node has a clustering label from ground truth knowledge for a network with community attributes. Here we focus on networks with non-overlapping communities. We define the community mapping function as $C_F$. It comes in two forms, the ground truth community knowledge $C_G$ and the result of community detection algorithms $C_A$. In a network with $K$ clusters, $C=\{C_1,C_2,\cdots,C_K\}$ denotes the set of clustering labels. Both $C_G$ and $C_A$ can be interpreted as functions satisfying $C_G,C_A : V \mapsto C$, namely $\forall v\in V$, it has its ground truth community attribute $C_G(v)\in C$ and algorithm result $C_A(v)\in C$. Our goal is to find an optimal link prediction algorithm that approaches $C_A$ as close to $C_G$ as possible. The closeness between $C_A$ and $C_G$ can be quantified nicely, this will be discussed later in chapter \ref{sec:expriment}.

For future reference, the main notations in this paper will be introduced here. $N_{C_i}$ stands for the set of nodes with the community attribute $C_i$ and $\|\cdot\|$ stands for taking the cardinality of the given set. $CM$ is the connection matrix satisfying $CM \in \mathbb{Z}^{K\times K}$. We let $CM(i,j)$ be the number of edges between cluster $C_i$ and $C_j$. Given the ground truth community $C_{G}$ and algorithm's output $C_{A}$, edge $e=(v_i,v_j)$ is a \emph{revising} edges if $C_{G}(v_i) = C_{G}(v_j)$ and $C_{A}(v_i) \neq C_{A}(v_j)$. At the same time we provide the definition of \emph{reinforcing} edges: edge $e=(v_i,v_j)$ is a reinforcing edge if and only if $C_{G}(v_i) = C_{G}(v_j)$ and $C_{A}(v_i) = C_{A}(v_j)$. These two definitions will be used to quantify the fixing power of different link prediction methods and visualize the transformation from revising process to reinforcing process.

\subsection{Inductive Biases}
It is worth noticing that in most cases, community detection methods might yield a larger number of clusters on both real-world and synthesized networks. Here we experiment with three representative community detection methods (LPA, Infomap and Louvain) to verify this phenomenon, since a fair amount of recently developed community detection methods are based on the intuitions behind these three methodologies \citep{okuda(2019)community, luo(2020)highly, roy(2021)nesifc}.

Here we present the community detection results of the three algorithms on nine real-world network systems in figure \ref{nine dataset}. Notice the difference in cluster numbers between ground truth knowledge and algorithm results. Except for the Louvain method on the Eurosis network, community detection methods consistently output a significant more number of communities than ground truth. Especially in the Cora\_OS dataset, the results of LPA and Infomap are at least 25 times larger than the actual value. Furthermore, according to  \citep{Burgess(2016)consensus}, experimental results of these community detection methods consistently yield a higher number of communities on LFR \citep{Lancichinetti(2008)LFR} benchmark graphs. All real-world datasets mentioned here will be formally introduced in section \ref{sec:expriment}.

\begin{figure*}[ht]
	\centering
	\subfloat[Ground Truth]{\includegraphics[width=2.5in]{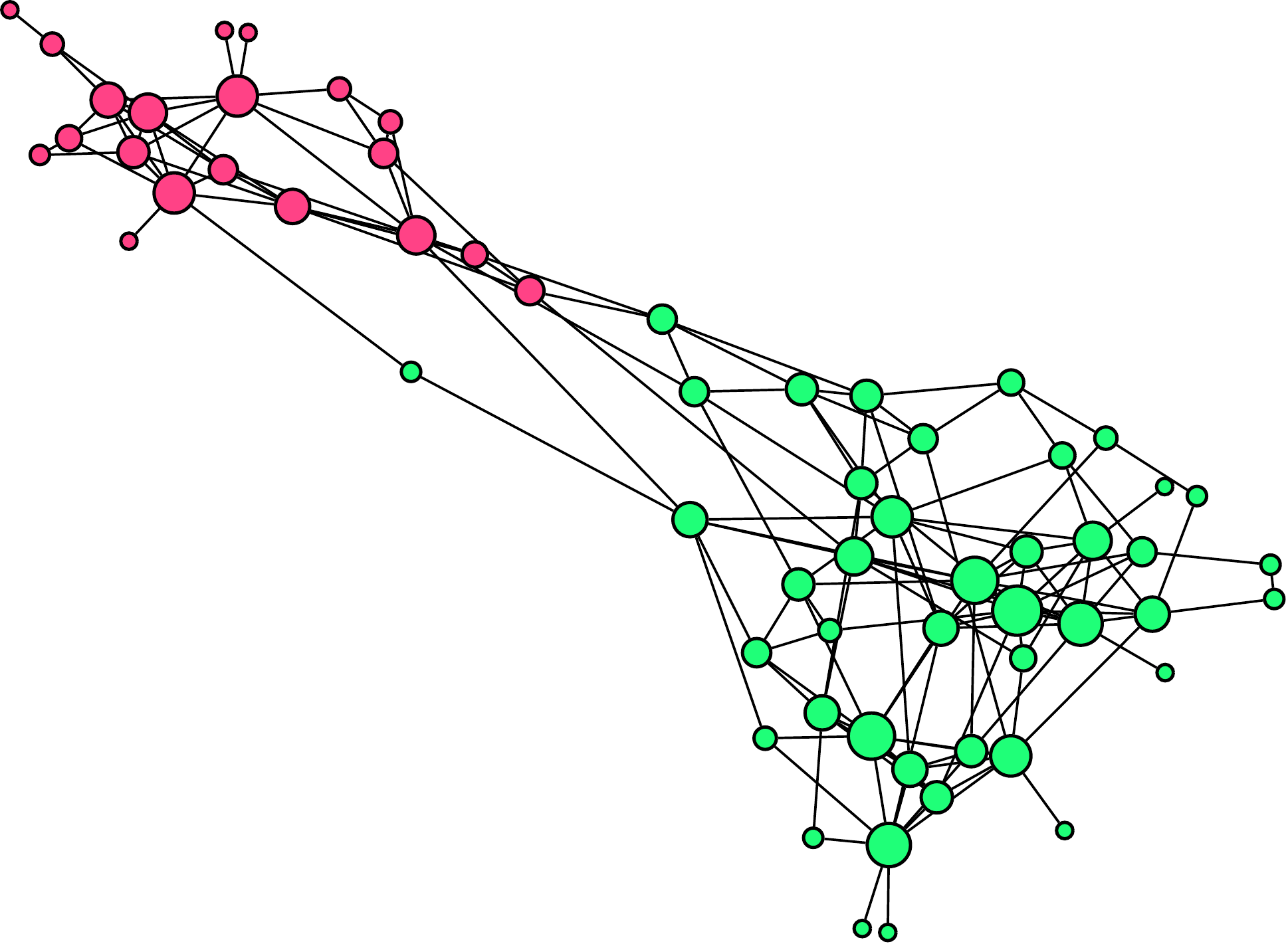}%
		\label{ground_truth}}
	\hfil
	\subfloat[Infomap Result]{\includegraphics[width=2.5in]{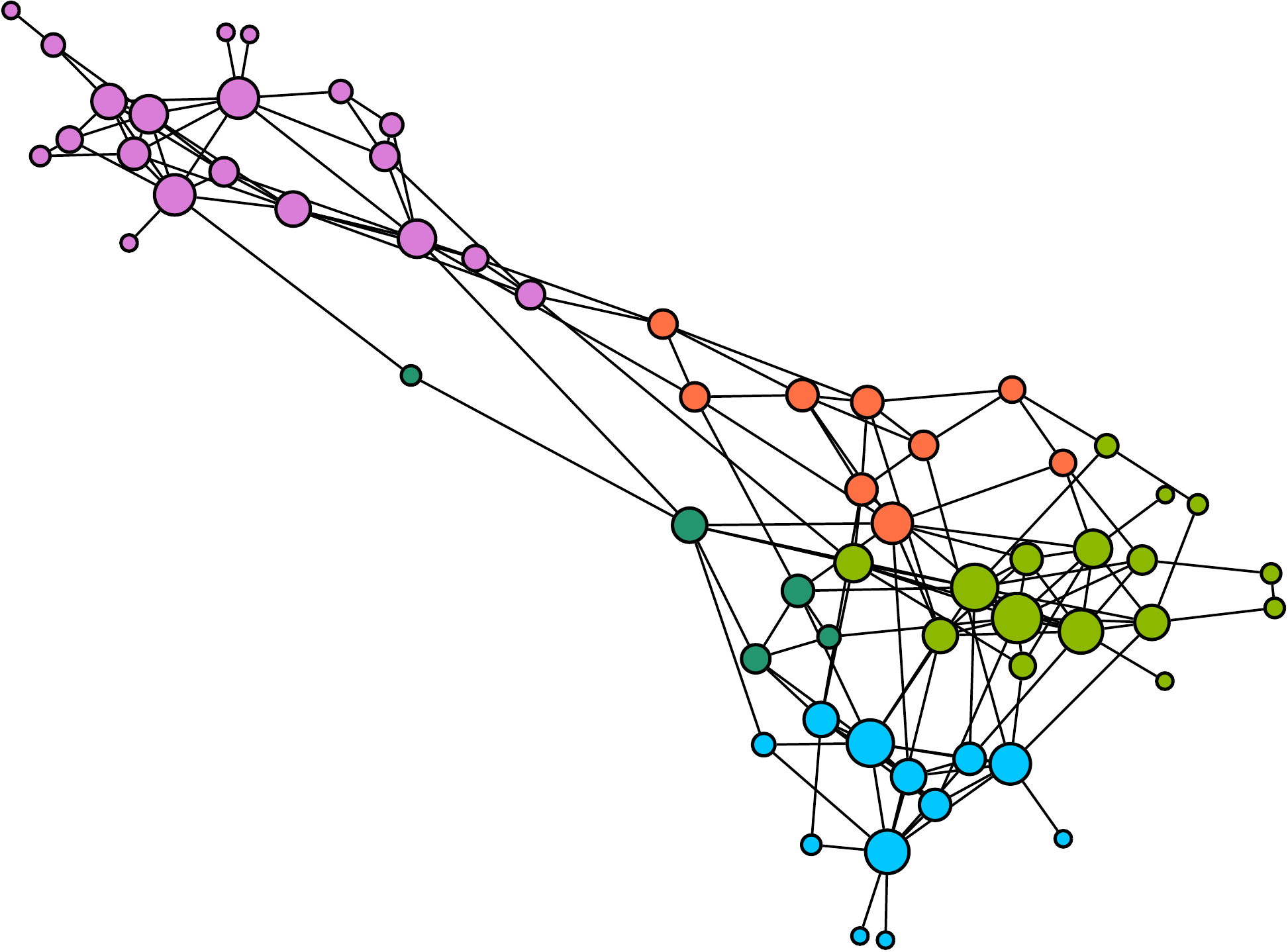}%
		\label{algorithm_output}}
	\caption{An example of the Dolphin network indicates that community detection algorithms turn a whole community into fragments. Compared with the ground truth information, the larger community in the Dolphin network gets fractured by the Infomap community detection method.}
	\label{dolphine contrast}
\end{figure*}

To take a step further, we can observe that the emergence of additional clusters comes from the fracture of complete ground truth communities due to the incompleteness of networks. For example, as illustrated in figure \ref{dolphine contrast}, the Dolphin network's larger ground truth community gets fragmented into four smaller communities by the Infomap method. Supposing the link prediction method could make connections among different parts of fractured subcommunities, community detection methods might achieve better performance by recognising and merging those subcommunities into a complete one.

If we could adjust the network topology structure by connecting nonexistent revising edges, the community detection algorithms would have a better chance to approach the ground truth community of networks. 

In conclusion, the inductive biases in this paper are listed as follows.
\begin{itemize}
	\item The outputs of community detection algorithms are highly likely to be incorrect and contain more clusters than actual cases because detection methods tend to split large communities into smaller ones.
	\item The connection of revising edges enhances the ambiguous community structure, which could help the downstream community detection algorithms perform better.
\end{itemize}

\subsection{Algorithm Skeleton}
It is commonly accepted that ambiguous community structure is challenging for community detection studies due to the subtle difference between inter-edges and intra-edges \citep{Su(2019)enhance}. The leading thought of the HAP link prediction method is to add links among fractured components of a complete community, thus turning misunderstood inter-edges into affirmative intra-edges.

As a community attribute related unsupervised link prediction method, the HAP method requires a community detection algorithm to trigger the community enhancement procedure.

From a general perspective, the proposed community structure enhancement method contains three main processes in each iteration: 1) community detection, 2) central and boundary nodes recognition, and 3) adding links. At the beginning of each iteration, an early-stopping criterion detects the procedure of community enhancement to determine the stopping point. The rest of this chapter mainly focuses on detailed information about each step. However, the community detection process will not be discussed here since it can be arbitrary methods given by the users.

\subsection{Central Nodes Recognition}
In order to achieve the goal of community enhancement, the HAP link prediction method first recognises central and boundary nodes. It then adds links to the network according to this information. The proposed community enhancement method has an evolutionary process which will be explained on short notice.

In central node recognition, most methods use centrality measures to define whether nodes are on the edge of communities or not  \citep{Su(2019)enhance, zhou(2021)robustecd}. Such centrality measures can be defined through the average distance between intra-community nodes, which can be regarded as geometric distance centrality. Other methods, such as calculating the fraction of neighbour nodes with the same community attributes, could be regarded as the probability of a one-step random walk ending within the community.

These two measures are both successful in identifying boundary nodes. However, the distance-based centrality measure will bring unwanted calculation complexity. Moreover, nodes with a larger degree are more likely to yield a smaller average distance. In the meantime, the node's neighbours with different community attributes are not fully considered. For one-step random walk measure, it has less computation complexity and takes the neighbour nodes' community attributes into consideration. However, the linear fractional expression of connection might not utilize the neighbourhood information fully since the diversity of communities is not considered.

\begin{figure}[h]
	\centering 
	\setlength{\belowcaptionskip}{-1cm}
	\includegraphics[width=0.34\textwidth]{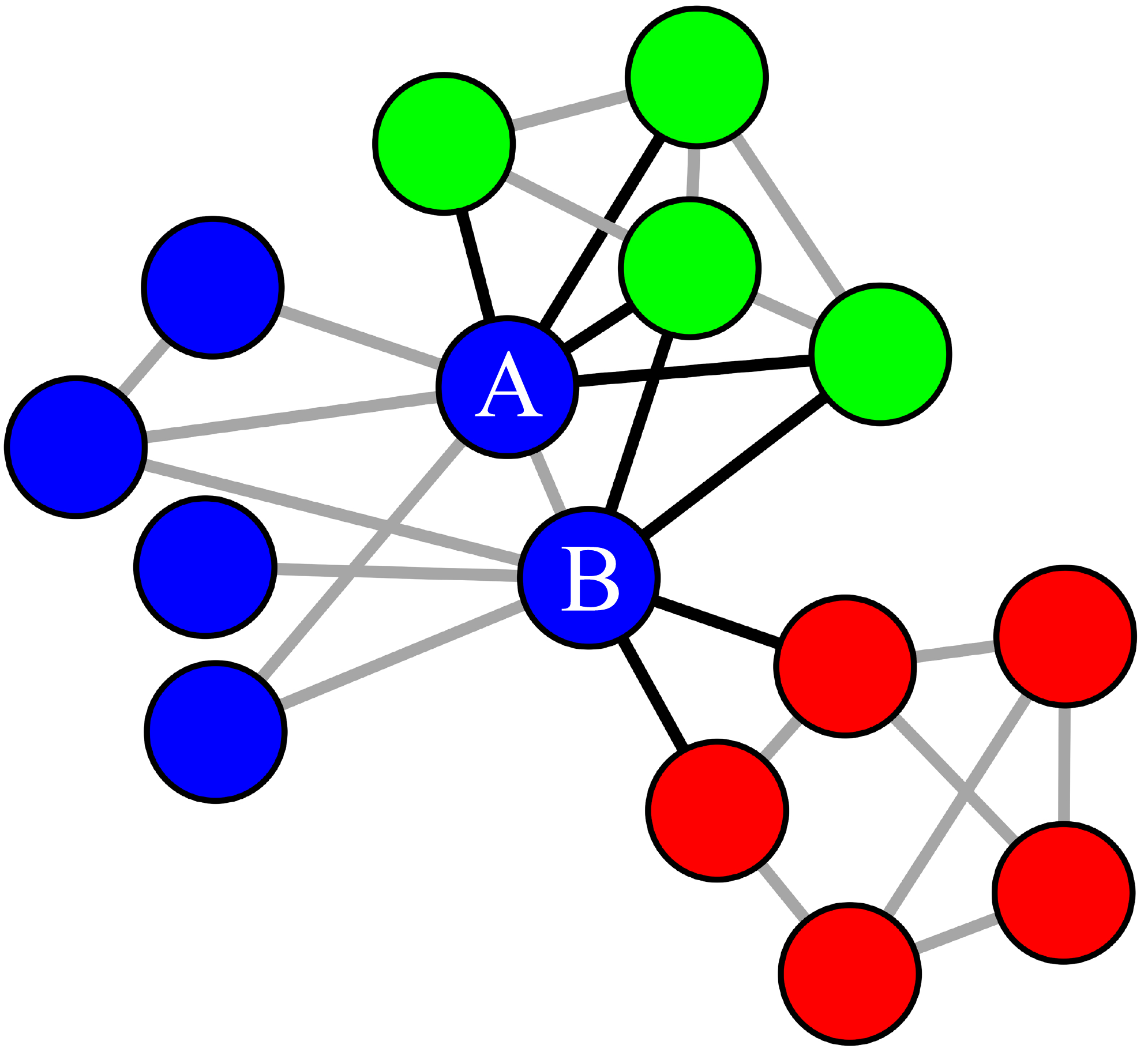}%
	\caption{Diagram of nodes A and B with their neighbourhood information. The diversity of nodes' colour indicates they belong to different communities.} 
	\label{centrality}
	
\end{figure}

For the distance-based centrality measure, the centrality score for a node $u$ with community attribute of $C_i$ is defined as:
\begin{equation}
	CS^{dis}_u = \frac{\sum_{x \in N_{C_i}}distance(u, x)}{\|N_{C_i}\|}. \label{equ:CS_dis}
\end{equation}
Take figure \ref{centrality} as an example, the distance-based centrality scores of nodes A and B equal $6/5$. For one-step random walk measure, in \citep{Su(2019)enhance}, the centrality score of a node $u$ with community attribute of $C_i$ is defined as: 
\begin{equation}
	CS^{rdm}_u = \frac{\|\Gamma(u)\cap N_{C_i}\|}{\|\Gamma(u)\|}  \label{equ:CS_u},
\end{equation}
where $\Gamma(u)$ stands for the neighbour of node $u$. The numerator stands for the number of nodes with the same community attribute while the denominator is the degree of node $u$, the one-hop random walk based centrality scores of nodes A and B equal $4/8$. For the above two measures, the centrality scores of nodes A and B are the same, but situations between nodes A and B are not identical where node A stands between two communities, while node B is on the overlapping section among three communities. To avoid this shortcoming, we should take the neighbour nodes' community attributes into deeper consideration.

Undoubtedly, the most helpful and easily accessible information for centrality measurement is the neighbourhood of nodes. To fully record such knowledge, we define the \emph{neighbourhood community enumeration (NCE)} of node $x$ as a multiset. Given the community mapping function $C_F$ and all its neighbour nodes $\Gamma(x)$:
\begin{equation}
	NCE_F(x)=\{C_F(i) \vert i \in \Gamma(x)\}  \label{equ:NCE}.
\end{equation}

Such as in figure \ref{centrality}, the NCEs of node A and B are $\{1,1,1,1,2,2,2,2\}$ ($1,2$ and $3$ represent for blue, green and red community respectively) and $\{1,1,1,1,2,2,3,3\}$ respectively. We apply the standard normalized Shannon entropy \citep{Shannon(1948)entropy} to quantify the uncertainty of NCE. If a node has complex neighbourhood information, namely a large entropy value, it is more likely to be a boundary node of a community. For a node $x$, given the community mapping function $C_F$, its \emph{boundary score (BS)} can be calculated as follow:
\begin{equation}
	BS(x) = \frac{H(NCE_F(x))}{log(\|\Gamma(x)\|)} \label{equ:BS}.
\end{equation}
In this equation, \emph{H} is the Shannon Entropy function and if $\|\Gamma(x)\| = 1$, its boundary score is set to $0$. It can be easily proven that the maximum value of the numerator is $log(\|\Gamma(x)\|)$. Thus the BS value always lies within interval $[0,1]$. For example, in figure \ref{centrality}, node A has a boundary score of $0.33$, and the boundary score of node B equals $0.50$. With the help of Boundary Score, the definition of \emph{consistency score (CS)} is defined as: 
\begin{equation}
	CS(x) = 1 - BS(x).
\end{equation}

The maximum value of CS of a node will be achieved when there is only one type of community in its neighbourhood. A higher CS value of a node suggests its neighbours' community attributes perform less uncertainty. This consistency measure can be viewed as a node centrality index evaluating the uncertainty of its neighbourhood clustering information.

\subsection{Link Connection}
\subsubsection{Harmony Similarity Measure}
Due to the imperfection of community detection algorithms, link prediction methods should not take the algorithm results of community detection as the ground truth knowledge. Experimental results in section \ref{sec:expriment} validate that improper use of community attributes would skew the community enhancement.

At first, we provide the similarity function for the RA index \citep{Zhou(2009)RA}:
\begin{equation}
	\label{equ:RA}
	s(x,y) = \sum_{z\in \Gamma(x)\cap\Gamma(y)}^{} \frac{1}{\|\Gamma(z)\|}.
\end{equation}
Equation \ref{equ:RA} is a measure to quantify the amount of information flow from node $x$ that is received by node $y$.

Following this train of thought, in the HAP method, the consideration of information spreading is combined with community attributes. Instead of spreading to neighbours uniformly, the information flow prefers to transmit information within the community through reliable central nodes. That is to say, the information flow is not point-to-point but in a community-to-community pattern.

Here we define the \emph{harmony (HM)} value to evaluate consistency between node $x,y$, namely:
\begin{equation}
	\label{equ:harmony}
	HM(x,y) = \frac{1}{\|\Gamma(x)\cap\Gamma(y)\|}\sum_{z\in \Gamma(x)\cap\Gamma(y)}^{} CS(z),
\end{equation}
where CS is the consistency score. Through equation \ref{equ:harmony}, it is clearly seen that the HM index is a second-order similarity measure via entropy. The HM score of two nodes is higher if their common neighbours have a larger average CS value. Following the definition of consistency, a node with a higher consistency value indicates that it is more likely to be a central node of a community. That is to say, node pair $x,y$ will yield a large harmony value if they share a neighbourhood of central nodes, suggesting that $x$ and $y$ are more likely to belong to the same community.

\subsubsection{Evolution Transformation}
According to equation \ref{equ:harmony}, nodes in the same community are expected to have higher HM scores. It is a desirable characteristic, but it tends to enhance the fractured communities instead of merging them at the beginning. Therefore this enhancement ability of link prediction is still insufficient. If the community enhancement procedure remains in the reinforcing process, it will consistently solidify small communities. This phenomenon will cause the community detection to output the same result, which is undesirable. 


In order to initiate the revising process and automatically transform it into the reinforcing procedure, the variables in such a process must be carefully designed. Following the demonstration of figure \ref{diagram}, we could imagine that the minimum size of communities will get larger in revising process. Furthermore, in figure \ref{dolphine contrast}, we can observe that the connection among fractured components from a large ground truth community is more frequent than different communities in the ground truth. Combining these two factors, we can infer that merging small communities should have higher priority at the beginning. Furthermore, for nodes $x,y$ from different communities, the HAP method should tend to add a link between them if their communities connect at a relatively high frequency.

Here we define \emph{Community Size Attribute (CSA)} with hub-preferred strategy:

\begin{equation}
	\label{equ:CSA}
	{CSA(x,y)} = \begin{cases}
		\frac{CM(C_x,C_y)}{\min \{CM(C_x,C_x), CM(C_y,C_y)\}},&{C_x \neq C_y}\\ 
		{\frac{\sqrt{\|N_{C_x}\|}}{\max_{i} \|N_{C_i}\|}},&{C_x = C_y} 
	\end{cases}
\end{equation}
CM is the connection matrix. Equation \ref{equ:CSA} calculates the ratio between inter-edges and intra-edges for smaller communities when $C_x \neq C_y$. When $C_x = C_y$, we do not want to enhance the community when its size is relatively small as aforementioned. Here we empirically set $CSA = \sqrt{N_{C_x}}/\max_{i}\{N_{C_i}\}$ in this situation, which performs satisfactorily in experiments. During the revising process, the minimum size of communities gets larger, leading to the CSA value for $C_x \neq C_y$ getting smaller, thus transforming into the reinforcing process.

Generally, the CSA index can be regarded as an indicator of community connection possibility. In contrast, the HM index explicitly points out the detail about which pair of nodes should build a connection. Combining these two indexes could achieve the community enhancement measure with the desired automatic transformation of such an evolutionary procedure. The similarity function of the HAP method is listed below:

\begin{equation}
	\label{HAP}
	HAP(x,y) = HM(x,y)\cdot CSA(x,y).
\end{equation}
\begin{algorithm}
	\caption{HAP Method}\label{algo:HAP}
	\begin{algorithmic}[1]
		\Require Graph $G=(V,E)$, Community Mapping Function $C_F$ and Number of Edge Adding $L$.
		\Ensure New Graph $G=(V,E')$.
		\State $NonExistEdges(NEE) \longleftarrow \emptyset$
		\For{$i\in V$}
			\For{$j\in \Gamma(i)$}
			\If{$(i,j) \notin E$}
			\State $NEE \longleftarrow NEE\cup (i,j)$
			\EndIf
			\EndFor
		\EndFor
		\State $HAP\ Scores \longleftarrow \emptyset$
		\State $CM \longleftarrow$ Conduct CM matrix by $C_F$
		\For{$e=(i,j)\in NEE$}
			\State $HM \longleftarrow$ Calculate HM value by eq. \ref{equ:harmony}
			\State $CSA \longleftarrow$ Calculate CSA value by eq. \ref{equ:CSA}
			\State $HAP \gets HM\cdot CSA$
			\State $HAP\ Scores \longleftarrow HAP\ Scores \cup (HAP,\ e)$
		\EndFor
		\State $HAP\ Scores \longleftarrow$ Descend Order($HAP\ Scores$)
		\State $E' \longleftarrow E\ \cup$ Top $L$ Edges In$(HAP\ Scores)$
		\State \Return $G = (V,E')$
	\end{algorithmic}
\end{algorithm}

\subsection{Early-stopping Criterion}
In order to prevent a trivial solution that all pairs of nodes are connected, resulting in a gigantic community, we here present an early-stopping criterion for long-term stability for the outputs of community detection methods. 

We need to determine when the community detection result stops improving to achieve this goal.  Since no prior knowledge about the ground truth is available, we have to employ an indicator solely based on community detection methods' output. The intuition is to determine when the enhancement measure steps into the reinforcing stage. At this stage, the result of community detection methods remains relatively stable.

Therefore, we propose an early-stopping method to calculate the NMI value between the latest and previous results. This method requires two hyperparameters, rounds of consideration $R$ and threshold value $\delta$. To be specific, assume that the community detection method has yielded $N$ results, denoted as $\{C_1, ..., C_N\}$, the early-stopping method calculates a set of NMI $\{NMI(C_{N-R-1}, C_N), ..., NMI(C_{N-1}, C_N)\}$, if the minimum value of this set is larger than $\delta$, the iteration process will be stopped immediately.

\subsection{Time Complexity Analysis}
For notations, we assume that $n$ is the number of nodes and $k$ is the largest degree of nodes. As mentioned earlier, the HAP mainly concerns about two-hop neighbours, so its time complexity is $O(nk^2*f(k)+g)$ where $f(k)$ is the computational cost to determine the likelihood for one pair of nodes and $g$ stands for the time complexity of preparation.

According to equation \ref{HAP}, $f(k)$ consists of calculating both CSA and HM values. For the CSA value, the time complexity of lookup is $O(1)$ once we construct the CM matrix. Determining the CM matrix requires going through all the links in a graph for one time, which has a time complexity of $O(n*k)$. In addition, it takes $O(n)$ to determine all the $|N_{C_i}|$ values. Both of the time complexity belongs to the term $g$. For calculation of HM value, according to equation \ref{equ:harmony}, each CS value is determined by traversing one-hop neighbour nodes for at most $k+k$ nodes. Combined, the time complexity of HM equals $O((k+k)*k)$, which belongs to the term $f(k)$.

That is to say, the time complexity of HAP is $O(nk^2*k^2+n*k+k)=O(nk^4)$. According to  \citep{martinez(2016)survey}, the computation cost of our method is competitive against all local methods since $k$ can be regarded as a constant due to the sparsity and incompleteness of networks. Furthermore, most community detection methods have much larger time complexity, confirming that our enhancement measure can be implemented as a plug-in module.

\begin{algorithm}
	\caption{Community Enhancement Method}\label{algo:CEM}
	\begin{algorithmic}[1]
		\Require Graph $G=(V,E)$, Community Detection Method $CDM$, Number of Edge Adding $L$, Maximun number of Interation $N$.
		\Ensure Community Mapping Function $C_F$, New Graph $G_{new}$.
		\State $G_{old} \longleftarrow G$
		\State $C_{old} \longleftarrow CDM(G_{old})$
		\State $Counter \longleftarrow 0$
		\While{$Counter \leq N$}
			\State $G_{new} \longleftarrow $HAP Method$(G_{old}, C_{old}, L)$
			\State $C_{new} \longleftarrow CDM(G_{new})$
			\State $G_{old} \longleftarrow G_{new}$
			\State $C_{old} \longleftarrow C_{new}$
			\If{Earlystopping}
			\State break
			\EndIf
			\State $Counter += 1$
		\EndWhile
		\State \Return $G = (V,E')$
	\end{algorithmic}
\end{algorithm}

With all the preparation work, the community enhancement process is demonstrated in algorithm \ref{algo:CEM}. In the experimental section, we will replace the HAP method with other link prediction methods to verify its validity.

\section{Experiments Results and Analysis}
\label{sec:expriment}
\subsection{Datasets}
Twelve networks are tested in the experimental process, including a network consisting of 62 dolphins in a community living off Doubtful Sound, New Zealand (Dolphins, for short) \citep{Lusseau(2003)dolphins}, network of friendships among 34 members of Zachary's karate club (Karate, for short) \citep{Zachary(1977)karate}, books about US politics sold by the online bookseller Amazon.com in 2004 (Polbooks, for short) \citep{Newman(2006)polbooks}, an interaction map between Science in Society actors on the Web of 12 European countries (Eurosis, for short), an online hyperlinks network between weblogs on US politics (Polblogs, for short) \citep{Adamic(2005)polblogs}, a network of the relationship between publication and the corresponding word from a dictionary (Cora, for short) \citep{Sen(2008)cora_sub}. A series of citation networks on different subdomains are listed as Cora Artificial Intelligence (Cora\_AI, for short), Cora Human Computer Interaction (Cora\_HCI, for short) and Cora Operating Systems (Cora\_OS, for short)  \citep{Subelj(2013)cora_full}, and three subsets of the 20 newsgroups dataset comprise around 18000 newsgroups posts on 20 topics (News\_1, News\_2, News\_3, for short) \citep{Yen(2007)news}.
\footnote{All codes and datasets are available on https://github.com/Young0313/HAP}

Details of the twelve datasets are listed in table \ref{dataset information}. The column of Transitivity indicates the fraction of close triangles in the network system, which is an indicator of connectivity. $N_C$ stands for the ground truth number of clusters in the network. The index of \textit{Intra} is defined as the ratio of edges within communities to demonstrate the community structure. \textit{L} is the hyperparameter of edge increment for each iteration, which is approximately proportional to the number of edges in the network system.

\begin{table}[h]
	\begin{center}
		\caption{Information of Twelve Real-world Networks\label{dataset information}}
		\centering
		\begin{tabular}{@{}lllllll@{}}
			\toprule
			Network  & \#Nodes & \#Edges & Transitivity & $N_C$ & Intra & $L$ \\ 
			\midrule
			Dolphins & 62    & 159   & 0.30878      & 2                   & 0.96226        & 10          \\
			Karate   & 34    & 78    & 0.25568      & 2                   & 0.87179        & 10          \\
			Polbooks & 105   & 441   & 0.34840      & 3                   & 0.84127        & 20          \\
			Eurosis  & 1272  & 6454  & 0.23478      & 13                  & 0.82290        & 100         \\
			Polblogs & 1222  & 16717 & 0.22596      & 2                   & 0.90578        & 200         \\
			Cora     & 2458  & 5069  & 0.09003      & 7                   & 0.80410        & 100         \\
			Cora\_AI & 4633  & 12985 & 0.15621      & 11                  & 0.82636        & 150         \\
			Cora\_HCI& 1053  & 2350  & 0.17730      & 5                   & 0.96213        & 50          \\
			Cora\_OS & 2068  & 8645  & 0.13664      & 4                   & 0.82852        & 120         \\
			News\_1  & 398   & 3347  & 0.42667      & 2                   & 0.93188        & 50          \\
			News\_2  & 598   & 5041  & 0.36420      & 3                   & 0.80401        & 100         \\
			News\_3  & 595   & 4557  & 0.35152      & 3                   & 0.85561        & 100         \\
			\botrule
		\end{tabular}
	\end{center}
	\setlength{\tabcolsep}{4mm}
\end{table}

\subsection{Baseline Methods and Evaluation Measures}
\subsubsection{Baseline Methods}
The kernel of the proposed community detection method is the likelihood-based link prediction algorithm. In order to verify the performance of proposed HAP method, it is compared with six other link prediction methods \citep{Burgess(2016)consensus}, including JA \citep{Jaccard(1901)JA}, PA \citep{Barabasi(1999)PA}, CN \citep{liben(2007)CN}, CN1 \citep{Soundarajan(2012)community}, RA \citep{Zhou(2009)RA} and RA1 \citep{Soundarajan(2012)community}. Detailed information about these methods can be found in these references. All seven link prediction methods are aggregated with the three community detection algorithms to get twenty-one combinations, which are tested on twelve real-world datasets. 

\subsubsection{Evaluation Measures}
In order to quantify the quality of community detection results, as well as measure the improvement of community enhancement, the widely used Normalized Mutual Information (NMI) evaluation methods will be introduced here:
$$NMI(X,Y)=\frac{MI(X,Y)}{\sqrt{H(X)\times H(Y)}}$$
where \emph{MI} \citep{Cover(1999)MI} stands for Mutual Information score and \emph{H} is the Shannon Entropy function. If we take $X$ as the algorithm output and $Y$ is the ground truth community structure, NMI=1 means perfect correlation while 0 stands for no mutual information. That is to say, higher NMI values mean better community detection results.

In addition to providing the best performance on NMI value, the improvement of NMI value will also be provided. The fraction of revising edges can quantify the correction ability of different link prediction methods. In order to verify the generalization ability and numerical stability, the difference between the original NMI value and the final output NMI value will be provided. Last, the dynamic transformation between the two stages will be demonstrated.

\subsection{Evaluation of NMI}

\begin{figure*}[h]
	\centering  
	\includegraphics[width=\textwidth]{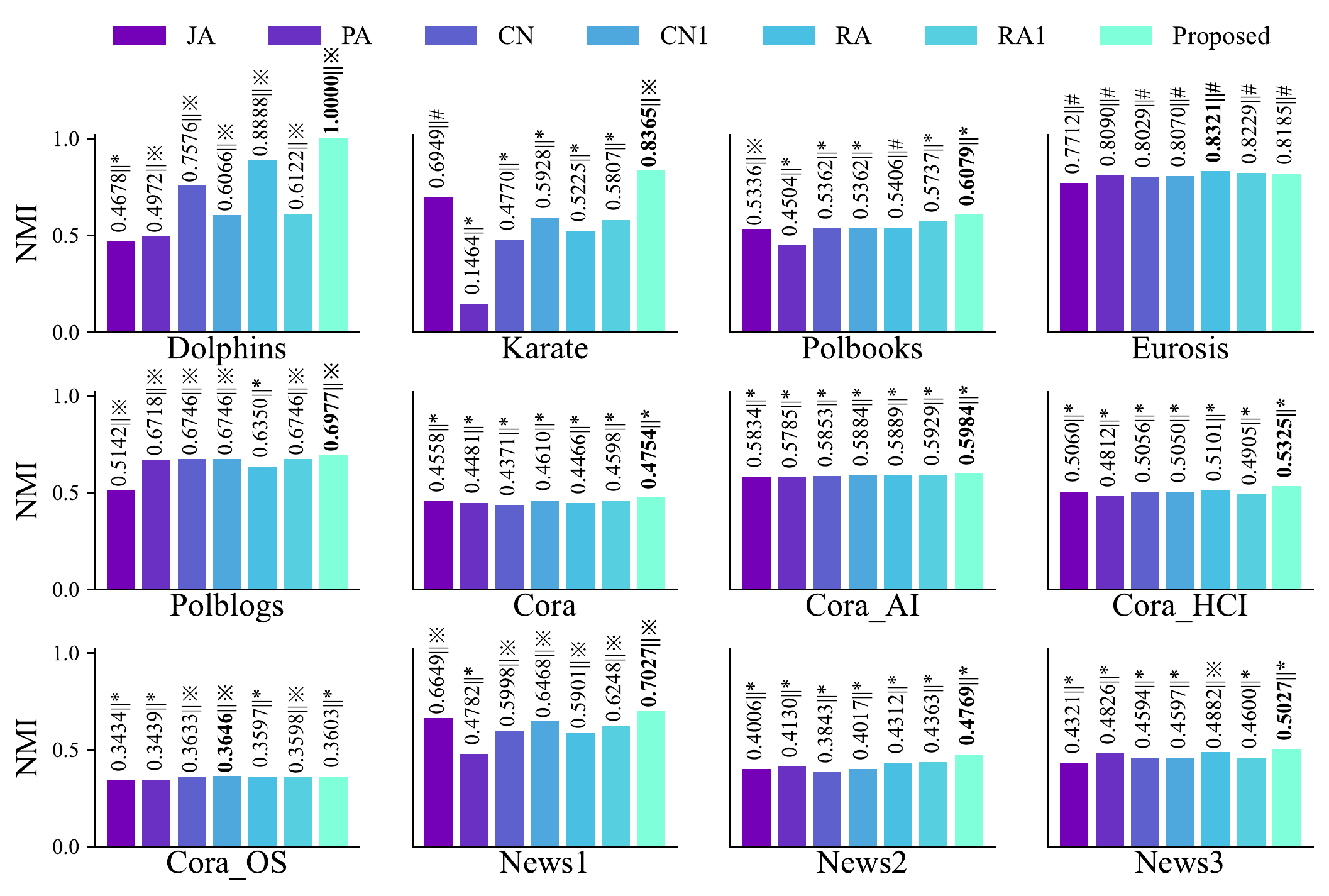}  
	\caption{Best NMI performance of all link prediction methods, labels after $||$ stand for the best performance achieved by Louvain($\ast$), Infomap($\#$) and LPA($\divideontimes$). Furthermore, we do not implement the Louvain algorithm on the Eurosis dataset because the experimental results in figure \ref{nine dataset} already demonstrated that the Lovain method achieved satisfying results, and the community enhancement task would not be significant.} 
	\label{best nmi}
\end{figure*}

Since the ground truth community structure behind those networks are all available, we compare the algorithm output with the ground truth by applying the NMI measure. The result of the best NMI performance of all link prediction methods among three different community detection algorithms is demonstrated in figure \ref{best nmi}.

\begin{table*}[h]
	\begin{center}
		\caption{NMI ranking of seven link prediction methods} 
		\label{nmi_ranking}
		\resizebox{\textwidth}{!}{
			\begin{tabular}{@{}cccccccccccccc@{}}
				\toprule
				& Dolphins   & Karate     & Polbooks   & Eurosis    & Polblogs   & Cora       & Cora\_AI   & Cora\_HCI  & Cora\_OS   & News\_1    & News\_2    & News\_3    & Mean \\ 
				\midrule
				JA       & 7          & 2          & 6          & 7          & 7          & 4          & 6          & 3          & 7          & 2          & 6          & 7          & 5.33 \\
				PA       & 6          & 7          & 7          & 4          & 5          & 5          & 7          & 7          & 6          & 7          & 4          & 3          & 5.67 \\
				CN       & 3          & 6          & 4          & 6          & 2          & 7          & 5          & 4          & 2          & 5          & 7          & 6          & 4.75 \\
				CN1      & 5          & 3          & 4          & 5          & 2          & 2          & 4          & 5          & \textbf{1} & 3          & 5          & 5          & 3.67 \\
				RA       & 2          & 5          & 3          & \textbf{1} & 6          & 6          & 3          & 6          & 5          & 6          & 3          & 2          & 3.67 \\
				RA1      & 4          & 4          & 2          & 2          & 2          & 3          & 2          & 2          & 4          & 4          & 2          & 4          & 3.25 \\
				HAP      & \textbf{1} &\textbf{1}  & \textbf{1} & 3          & \textbf{1} & \textbf{1} & \textbf{1} & \textbf{1} & 3          & \textbf{1} & \textbf{1} & \textbf{1} & 1.33 \\ 
				\botrule
		\end{tabular}}
	\end{center}
\end{table*}

For each network, we independently apply the twenty-one combinations of link prediction methods with community detection algorithms and repeat the link adding and clustering for ten iterations as the community enhancement. The results in figure \ref{best nmi} are the best performance for each link prediction method among the three community detection algorithms. As shown in table \ref{nmi_ranking}, our proposed HAP link prediction method has top performance in 10 out of 12 networks, especially in the Dolphin network it reaches the maximum value of NMI as 1.

\begin{figure*}[h]
	\centering  
	\includegraphics[width=\textwidth]{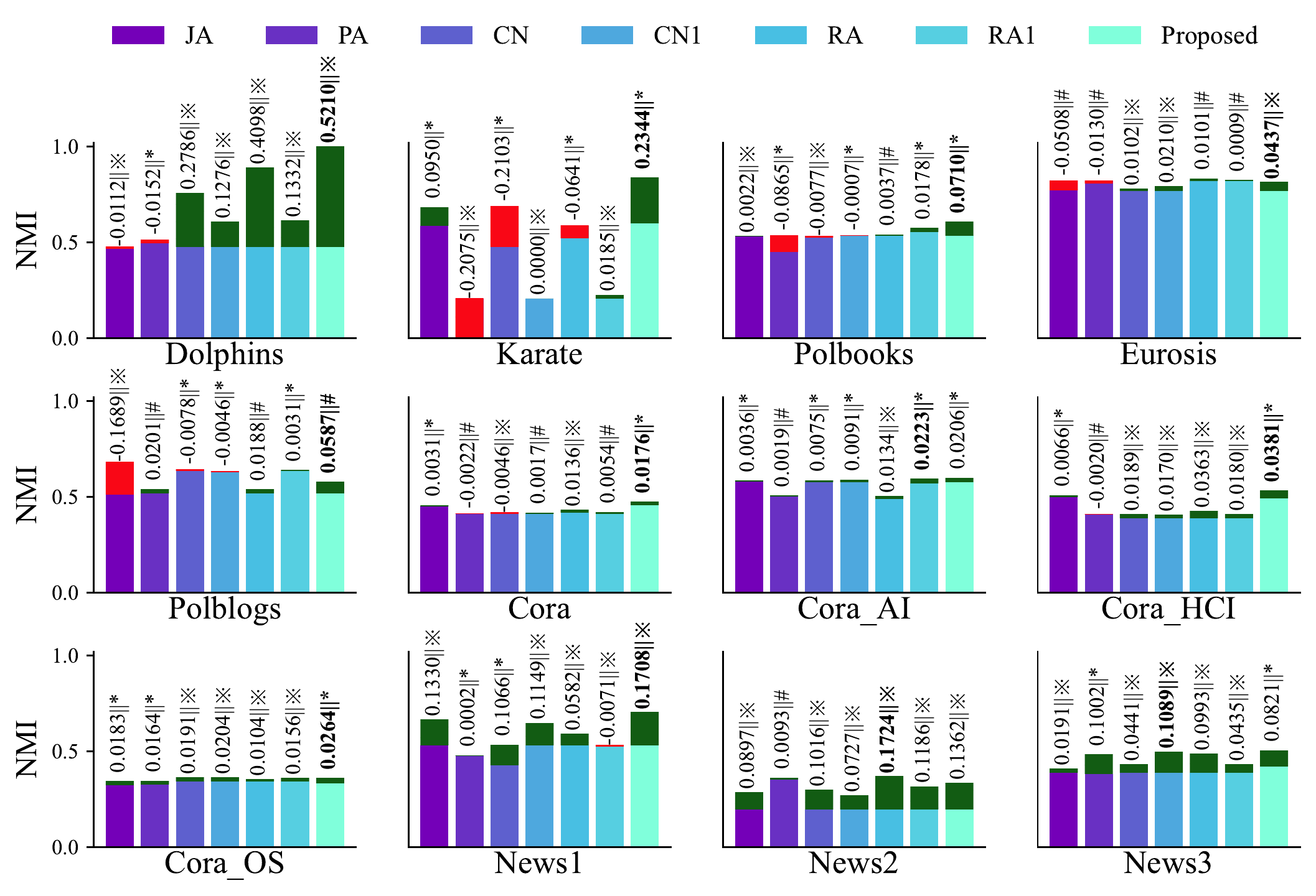}  
	\caption{Largest improvement for each combination, labels after $||$ stand for best performance achieved by Louvain($\ast$), Infomap($\#$) and LPA($\divideontimes$). The Louvain community detection on the Eurosis dataset is also excluded.} 
	\label{best improvement}
\end{figure*}

Not limited to the best performance of NMI values, the improvement on each real-world dataset is still of great importance. Since there remains uncertainty in the community detection section, especially in large-scale networks, several link prediction methods might get leading performance since they acquire better community detection results at the beginning of the community enhancement process (demonstrated in algorithm \ref{algo:CEM}). Here we provide such information in figure \ref{best improvement}, where the red part indicates the final result is decreased. In contrast, the green part shows the improvement between the initial partition result and the terminal output after the final round of community enhancement.

\begin{table*}[h]
	\begin{center}
		\caption{Best NMI improvement ranking of seven link prediction methods} 
		\label{nmi_improve_ranking}
		\resizebox{\textwidth}{!}{
			\begin{tabular}{@{}cccccccccccccc@{}}
				\hline
				& Dolphins   & Karate     & Polbooks   & Eurosis    & Polblogs   & Cora       & Cora\_AI   & Cora\_HCI  & Cora\_OS   & News\_1    & News\_2    & News\_3    & Mean \\ \hline
				JA       & 6          & 2          & 4          & 7          & 7          & 4          & 6          & 6          & 4          & 2          & 5          & 7          & 5.00 \\
				PA       & 7          & 6          & 7          & 6          & 2          & 6          & 7          & 7          & 5          & 6          & 7          & 2          & 5.67 \\
				CN       & 3          & 7          & 6          & 3          & 6          & 7          & 5          & 3          & 3          & 4          & 4          & 5          & 4.67 \\
				CN1      & 5          & 4          & 5          & 2          & 5          & 5          & 4          & 5          & 2          & 3          & 6          & \textbf{1} & 3.92 \\
				RA       & 2          & 5          & 3          & 4          & 3          & 2          & 3          & 2          & 7          & 5          & \textbf{1} & 3          & 3.33 \\
				RA1      & 4          & 3          & 2          & 5          & 4          & 3          & \textbf{1} & 4          & 6          & 7          & 3          & 6          & 4.00 \\
				HAP      & \textbf{1} &\textbf{1}  & \textbf{1} & \textbf{1} & \textbf{1} & \textbf{1} & 2          & \textbf{1} & \textbf{1} & \textbf{1} & 2          & 4          & 1.42 \\ \hline
		\end{tabular}}
	\end{center}
\end{table*}

From figure \ref{best improvement}, we could find that our proposed methods have top-2 performance in most cases (11 out of 12).  In addition, our proposed HAP method improves performance on all datasets. The corresponding ranking is demonstrated in table \ref{nmi_improve_ranking}. Furthermore, by comparing table \ref{nmi_ranking} with table \ref{nmi_improve_ranking}, we can verify our hypothesis such as the main contribution for RA1 measure's higher NMI values originates in better initial partition result rather than its community enhancing ability.

Furthermore, observing the average ranks of all baseline methods in table \ref{nmi_ranking} and \ref{nmi_improve_ranking}, we find that the RA family (RA, RA1) generally performs better than the CN family (CN, CN1). Considering the characteristics and intuitions behind those link prediction methods, we might conclude that the information flow paradigm (RA) might be more appropriate than the neighbourhood overlapping index (CN) in the scene of community enhancement in real-world data.

\subsection{Evaluation of Revising Edges}

Taking a step further to explore the reason for such leading performance by the HAP method, we need to consider the role of revising edges in the community enhancement task. According to inductive biases, it is believed that the connection of revising edges will profoundly affect the community detection methods and lead those algorithms to better NMI performance. Table \ref{Fraction} shows the revising ability of these seven link prediction methods from the perspective of what kind of edges they add to the networks. For each link prediction method, we will calculate the fraction between revising edges and all edges for all three community detection methods. Each value in this table is the average proportion of revising edges among three community detection methods. For example, if the fraction of revising edges for the CN index on three community detection methods are 40 out of 100, 30 out of 100 and 35 out of 100, then the corresponding value in this table is $(40/100 + 30/100 + 35/100)/3 *100\% = 35\%$. The best value on each network is highlighted.

\begin{table*}[h]
	\begin{center}
		\caption{Fraction of revising edges for all methods on 12 real-world networks(\%)} 
		\label{Fraction}
		\resizebox{\textwidth}{!}{
			\begin{tabular}{lllllllllllll}
				\hline
				\multicolumn{1}{c}{Methods} & \multicolumn{1}{c}{Dolphins} & \multicolumn{1}{c}{Karate} & \multicolumn{1}{c}{Polbooks} & \multicolumn{1}{c}{Eurosis} & \multicolumn{1}{c}{Polblogs} & \multicolumn{1}{c}{Cora} & \multicolumn{1}{c}{Cora\_AI} & \multicolumn{1}{c}{Cora\_HCI} & \multicolumn{1}{c}{Cora\_OS} & \multicolumn{1}{c}{News\_1} & \multicolumn{1}{c}{News\_2} & \multicolumn{1}{c}{News\_3} \\ \hline
				JA                          & 6.33                         & 9.00                       & 5.67                         & 3.60                        & 1.58                         & 5.87                     & 4.47                         & 6.00                          & 7.43                         & 1.67                        & 4.10                        & 6.47                        \\
				PA                          & 15.33                        & 6.33                       & 6.83                         & 0.70                        & 0.95                         & 5.47                     & 7.93                         & 43.73                         & 20.90                        & 23.93                       & 14.27                       & 14.07                       \\
				CN                          & 18.67                        & 7.33                       & 0.00                         & 0.35                        & 0.05                         & 3.47                     & 1.56                         & 4.60                          & 9.30                         & 2.93                        & 5.43                        & 13.43                       \\
				CN1                         & 4.33                         & 4.00                       & 0.00                         & 0.00                        & 0.00                         & 0.13                     & 0.07                         & 0.20                          & 1.10                         & 0.00                        & 0.00                        & 0.33                        \\
				RA                          & 32.33                        & 12.00                      & 5.67                         & 5.60                        & 1.00                         & 23.73                    & 19.64                        & 21.95                         & 23.90                        & 11.80                       & 12.30                       & 14.93                       \\
				RA1                         & 12.67                        & 8.00                       & 0.50                         & 0.40                        & 0.13                         & 2.90                     & 1.29                         & 2.07                          & 2.93                         & 0.00                        & 0.07                        & 0.17                        \\
				HAP                         & \textbf{82.67}               & \textbf{21.00}             & \textbf{59.33}               & \textbf{53.20}              & \textbf{54.25}               & \textbf{78.33}           & \textbf{74.20}               & \textbf{90.00}                & \textbf{88.47}               & \textbf{87.67}              & \textbf{57.67}              & \textbf{48.57}              \\ \hline
		\end{tabular}}
	\end{center}
\end{table*}

\begin{figure*}[htp]
	\centering  
	\includegraphics[width=\textwidth]{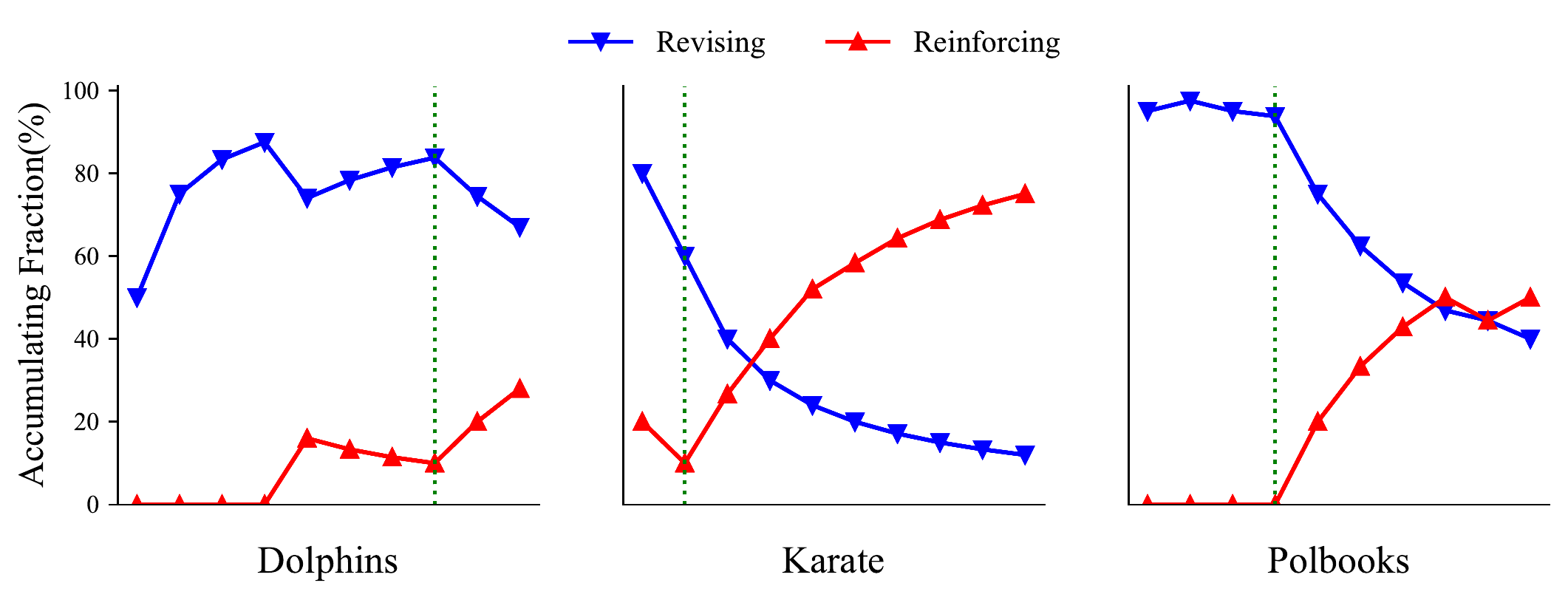}  
	\caption{The accumulating fraction of revising edges and reinforcing edges, the difference between two consecutive points indicates the incremental of the corresponding type of edges, and the green vertical dash line unveils the auto transformation between two processes.} 
	\label{transformation}
\end{figure*}

As can be seen from table \ref{Fraction}, the proposed algorithm can achieve the best performance in all datasets with the overwhelmingly highest fraction of revising edges. Additionally, comparing the CN and RA index with their community attributes version CN1 and RA1, the experimental results on revising fraction elucidate that improper consideration of community attributes will damage the enhancement measure's correction ability.

\begin{figure*}[htp]
	\centering
	\subfloat[Stage 1]{\includegraphics[width=0.25\textwidth]{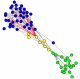}%
		\label{dolphine_stage_1}}
	\hfil
	\subfloat[Intermediate]{\includegraphics[width=0.25\textwidth]{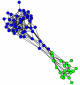}%
		\label{dolphine_middle}}
	\hfil
	\subfloat[Stage 2]{\includegraphics[width=0.25\textwidth]{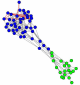}%
		\label{dolphine_stage_2}}
	\caption{This figure demonstrates three snapshots of the experimental procedure of the Dolphin network. The red line indicates the links that are appended in the corresponding round. Subfigure \ref{dolphine_stage_1} to subfigure \ref{dolphine_middle} represents the revising stage of community enhancement, which makes connection between fractured components and turns them into a complete community. While subfigure \ref{dolphine_middle} to subfigure \ref{dolphine_stage_2} shows the reinforcing process to enhance the correct community detection result.} 
	\label{dolphine experiment}
\end{figure*}

Revising and reinforcing edges both play an essential role in mending fractured communities. Thus we design our HAP method with an evolution process to transform from adding revising edges to adding reinforcing edges. The transformation has been successfully captured in several datasets. As demonstrated in figure \ref{transformation}, the community enhancement method can automatically transform the revising process into reinforcing. The illustration of the transformation of the HAP method on the Dolphin network is presented in figure \ref{dolphine experiment}, where the evolutionary process between two stages is captured.

\begin{table*}[h]
	\centering
	\renewcommand\arraystretch{1.5} 
	\caption{Experimental result of 12 real-world datasets on twenty-one combinations of link prediction methods and community detection methods. Sings in this table show the difference between original and final NMI values.}
	\label{Full Experiment}
	\resizebox{\textwidth}{!}{
		\begin{tabular}{cccccccccccccc}
			\hline
			LP                    & CD      & Dolphins                          & Karate                            & Polbooks                          & Eurosis                           & Polblogs                          & Cora                              & Cora\_AI                          & Cora\_HCI                         & Cora\_OS                          & News\_1                           & News\_2                           & News\_3                           \\ \hline
			& Louvain & -                                 & {\color[HTML]{FF0000} \textbf{+}} & -                                 & \textbackslash{}                  & -                                 & {\color[HTML]{FF0000} \textbf{+}} & {\color[HTML]{FF0000} \textbf{+}} & {\color[HTML]{FF0000} \textbf{+}} & {\color[HTML]{FF0000} \textbf{+}} & -                                 & -                                 & -                                 \\
			& Infomap & -                                 & -                                 & -                                 & -                                 & -                                 & -                                 & -                                 & -                                 & {\color[HTML]{FF0000} \textbf{+}} & -                                 & {\color[HTML]{FF0000} \textbf{+}} & -                                 \\
			\multirow{-3}{*}{JA}  & LPA     & -                                 & {\color[HTML]{FF0000} \textbf{+}} & -                                 & -                                 & -                                 & -                                 & -                                 & -                                 & -                                 & {\color[HTML]{FF0000} \textbf{+}} & {\color[HTML]{FF0000} \textbf{+}} & {\color[HTML]{FF0000} \textbf{+}} \\ \hline
			& Louvain & -                                 & -                                 & -                                 & \textbackslash{}                  & {\color[HTML]{FF0000} \textbf{+}} & -                                 & -                                 & -                                 & {\color[HTML]{FF0000} \textbf{+}} & {\color[HTML]{FF0000} \textbf{+}} & -                                 & {\color[HTML]{FF0000} \textbf{+}} \\
			& Infomap & -                                 & -                                 & -                                 & -                                 & {\color[HTML]{FF0000} \textbf{+}} & -                                 & {\color[HTML]{FF0000} \textbf{+}} & -                                 & {\color[HTML]{FF0000} \textbf{+}} & -                                 & {\color[HTML]{FF0000} \textbf{+}} & -                                 \\
			\multirow{-3}{*}{RA}  & LPA     & -                                 & -                                 & -                                 & -                                 & -                                 & -                                 & -                                 & -                                 & -                                 & -                                 & -                                 & -                                 \\ \hline
			& Louvain & {\color[HTML]{FF0000} \textbf{+}} & -                                 & -                                 & \textbackslash{}                  & -                                 & -                                 & {\color[HTML]{FF0000} \textbf{+}} & -                                 & -                                 & {\color[HTML]{FF0000} \textbf{+}} & -                                 & {\color[HTML]{FF0000} \textbf{+}} \\
			& Infomap & {\color[HTML]{FF0000} \textbf{+}} & -                                 & -                                 & -                                 & -                                 & -                                 & -                                 & {\color[HTML]{FF0000} \textbf{+}} & -                                 & -                                 & {\color[HTML]{FF0000} \textbf{+}} & -                                 \\
			\multirow{-3}{*}{CN}  & LPA     & {\color[HTML]{FF0000} \textbf{+}} & -                                 & -                                 & {\color[HTML]{FF0000} \textbf{+}} & -                                 & -                                 & {\color[HTML]{FF0000} \textbf{+}} & {\color[HTML]{FF0000} \textbf{+}} & {\color[HTML]{FF0000} \textbf{+}} & {\color[HTML]{FF0000} \textbf{+}} & {\color[HTML]{FF0000} \textbf{+}} & {\color[HTML]{FF0000} \textbf{+}} \\ \hline
			& Louvain & {\color[HTML]{FF0000} \textbf{+}} & -                                 & -                                 & \textbackslash{}                  & -                                 & {\color[HTML]{FF0000} \textbf{+}} & {\color[HTML]{FF0000} \textbf{+}} & {\color[HTML]{FF0000} \textbf{+}} & -                                 & {\color[HTML]{FF0000} \textbf{+}} & -                                 & {\color[HTML]{FF0000} \textbf{+}} \\
			& Infomap & -                                 & -                                 & -                                 & -                                 & -                                 & {\color[HTML]{FF0000} \textbf{+}} & -                                 & {\color[HTML]{FF0000} \textbf{+}} & {\color[HTML]{FF0000} \textbf{+}} & -                                 & {\color[HTML]{FF0000} \textbf{+}} & -                                 \\
			\multirow{-3}{*}{CN1} & LPA     & {\color[HTML]{FF0000} \textbf{+}} & =                                 & -                                 & {\color[HTML]{FF0000} \textbf{+}} & -                                 & -                                 & {\color[HTML]{FF0000} \textbf{+}} & {\color[HTML]{FF0000} \textbf{+}} & {\color[HTML]{FF0000} \textbf{+}} & {\color[HTML]{FF0000} \textbf{+}} & {\color[HTML]{FF0000} \textbf{+}} & {\color[HTML]{FF0000} \textbf{+}} \\ \hline
			& Louvain & -                                 & -                                 & -                                 & \textbackslash{}                  & -                                 & {\color[HTML]{FF0000} \textbf{+}} & {\color[HTML]{FF0000} \textbf{+}} & {\color[HTML]{FF0000} \textbf{+}} & {\color[HTML]{FF0000} \textbf{+}} & -                                 & {\color[HTML]{FF0000} \textbf{+}} & -                                 \\
			& Infomap & -                                 & -                                 & {\color[HTML]{FF0000} \textbf{+}} & {\color[HTML]{FF0000} \textbf{+}} & {\color[HTML]{FF0000} \textbf{+}} & {\color[HTML]{FF0000} \textbf{+}} & {\color[HTML]{FF0000} \textbf{+}} & {\color[HTML]{FF0000} \textbf{+}} & {\color[HTML]{FF0000} \textbf{+}} & {\color[HTML]{FF0000} \textbf{+}} & -                                 & {\color[HTML]{FF0000} \textbf{+}} \\
			\multirow{-3}{*}{RA}  & LPA     & {\color[HTML]{FF0000} \textbf{+}} & -                                 & -                                 & {\color[HTML]{FF0000} \textbf{+}} & -                                 & {\color[HTML]{FF0000} \textbf{+}} & {\color[HTML]{FF0000} \textbf{+}} & {\color[HTML]{FF0000} \textbf{+}} & {\color[HTML]{FF0000} \textbf{+}} & {\color[HTML]{FF0000} \textbf{+}} & {\color[HTML]{FF0000} \textbf{+}} & {\color[HTML]{FF0000} \textbf{+}} \\ \hline
			& Louvain & -                                 & -                                 & {\color[HTML]{FF0000} \textbf{+}} & \textbackslash{}                  & {\color[HTML]{FF0000} \textbf{+}} & -                                 & {\color[HTML]{FF0000} \textbf{+}} & -                                 & -                                 & {\color[HTML]{FF0000} \textbf{+}} & -                                 & -                                 \\
			& Infomap & {\color[HTML]{FF0000} \textbf{+}} & -                                 & {\color[HTML]{FF0000} \textbf{+}} & {\color[HTML]{FF0000} \textbf{+}} & {\color[HTML]{FF0000} \textbf{+}} & {\color[HTML]{FF0000} \textbf{+}} & {\color[HTML]{FF0000} \textbf{+}} & {\color[HTML]{FF0000} \textbf{+}} & {\color[HTML]{FF0000} \textbf{+}} & -                                 & {\color[HTML]{FF0000} \textbf{+}} & -                                 \\
			\multirow{-3}{*}{RA1} & LPA     & {\color[HTML]{FF0000} \textbf{+}} & {\color[HTML]{FF0000} \textbf{+}} & -                                 & {\color[HTML]{FF0000} \textbf{+}} & -                                 & {\color[HTML]{FF0000} \textbf{+}} & {\color[HTML]{FF0000} \textbf{+}} & {\color[HTML]{FF0000} \textbf{+}} & {\color[HTML]{FF0000} \textbf{+}} & {\color[HTML]{FF0000} \textbf{+}} & {\color[HTML]{FF0000} \textbf{+}} & {\color[HTML]{FF0000} \textbf{+}} \\ \hline
			& Louvain & {\color[HTML]{FF0000} \textbf{+}} & {\color[HTML]{FF0000} \textbf{+}} & {\color[HTML]{FF0000} \textbf{+}} & \textbackslash{}                  & {\color[HTML]{FF0000} \textbf{+}} & {\color[HTML]{FF0000} \textbf{+}} & {\color[HTML]{FF0000} \textbf{+}} & {\color[HTML]{FF0000} \textbf{+}} & {\color[HTML]{FF0000} \textbf{+}} & {\color[HTML]{FF0000} \textbf{+}} & {\color[HTML]{FF0000} \textbf{+}} & {\color[HTML]{FF0000} \textbf{+}} \\
			& Infomap & {\color[HTML]{FF0000} \textbf{+}} & {\color[HTML]{FF0000} \textbf{+}} & -                                 & -                                 & {\color[HTML]{FF0000} \textbf{+}} & {\color[HTML]{FF0000} \textbf{+}} & {\color[HTML]{FF0000} \textbf{+}} & {\color[HTML]{FF0000} \textbf{+}} & {\color[HTML]{FF0000} \textbf{+}} & {\color[HTML]{FF0000} \textbf{+}} & {\color[HTML]{FF0000} \textbf{+}} & {\color[HTML]{FF0000} \textbf{+}} \\
			\multirow{-3}{*}{HAP} & LPA     & {\color[HTML]{FF0000} \textbf{+}} & {\color[HTML]{FF0000} \textbf{+}} & {\color[HTML]{FF0000} \textbf{+}} & {\color[HTML]{FF0000} \textbf{+}} & {\color[HTML]{FF0000} \textbf{+}} & {\color[HTML]{FF0000} \textbf{+}} & -                                 & {\color[HTML]{FF0000} \textbf{+}} & {\color[HTML]{FF0000} \textbf{+}} & {\color[HTML]{FF0000} \textbf{+}} & {\color[HTML]{FF0000} \textbf{+}} & {\color[HTML]{FF0000} \textbf{+}} \\ \hline
	\end{tabular}}
\end{table*}

Last but not least, table \ref{Full Experiment} shows the sign of the difference between the original NMI value and the final output NMI value. The improvement is labelled in red in this table with the sign $+$, while the sign $-$ indicates a decrease. It can be seen that our HAP method has the most significant application stability and suits 17 out of 20 cases.

\section{Conclusion and Future Work}
\label{sec:conclusion}
In this paper, we propose the HAP method to realize link prediction based network community enhancement methods. Combining the HM and CS indexes, it can be treated as an iterative method that first determines the connection from a global perspective (community level) and then focuses on the neighbourhood's local information (node level) at each iteration. It has desirable portability and simplicity with low computation cost. Compared with other baseline methods on real-world datasets, our proposed novel index performs better in most cases. Furthermore, the HAP method is unsupervised, which can be implemented in most scenarios. Finally, thanks to the iteration paradigm, all local link prediction approaches are no longer bounded by the two-hop distance. 

There remain some works that need to be further studied. In this paper, we only consider adding edges to the network system, but removing existing edges also needs to be considered. Moreover, the square root term in the similarity function of the proposed method is an empirical modification term which cannot suit all situations. In future work, we will further explore the preprocessing ability of link prediction methods in other tasks like node classification in real-world scenarios.

\bmhead{Acknowledgments}

This work is supported by the National Natural Science Foundation of China (Grant Nos. 62276013, 62141605, 62050132), the Beijing Natural Science Foundation (Grant No. 1192012), and the Fundamental Research Funds for the Central Universities.


\bibliography{mylib}

\vspace{1cm}
\setlength\intextsep{0pt}
\begin{wrapfigure}{l}{25mm} 
	\includegraphics[width=1in,height=1.25in,clip,keepaspectratio]{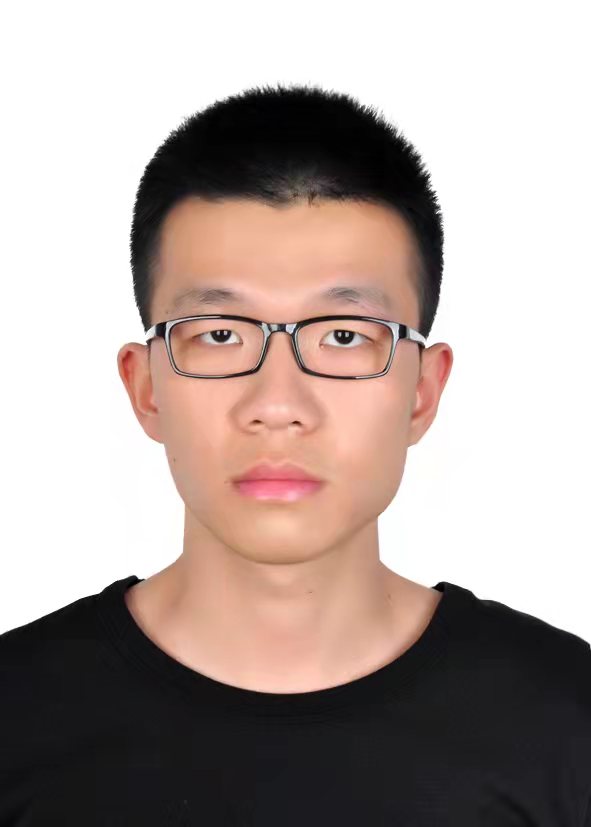}
\end{wrapfigure}\par
\textbf{Qiming Yang} received the B.S. degree from Beihang University, Beijing, China, in 2021. He is currently pursuing the master's degree with the School of Mathematical Sciences, Beihang University, Beijing, China. His current research interests include complex networks, link prediction and graph representation learning.

\subsection*{ }
\subsection*{ }
\setlength\intextsep{0pt}
\begin{wrapfigure}{l}{25mm} 
	\includegraphics[width=1in,height=1.25in,clip,keepaspectratio]{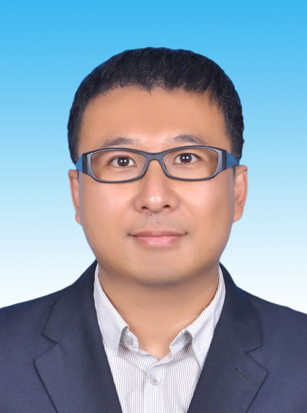}
\end{wrapfigure}\par
\textbf{Wei Wei} received the Ph.D. degree in mathematics from the School of Mathematical Sciences, Peking University, Beijing, China, in 2009. He is currently an Associate Professor with the School of Mathematical Sciences, Beihang University, Beijing, China. His research interests include graph learning, complexity and complex system.

\subsection*{ }
\subsection*{ }
\setlength\intextsep{0pt}
\begin{wrapfigure}{l}{25mm} 
	\includegraphics[width=1in,height=1.25in,clip,keepaspectratio]{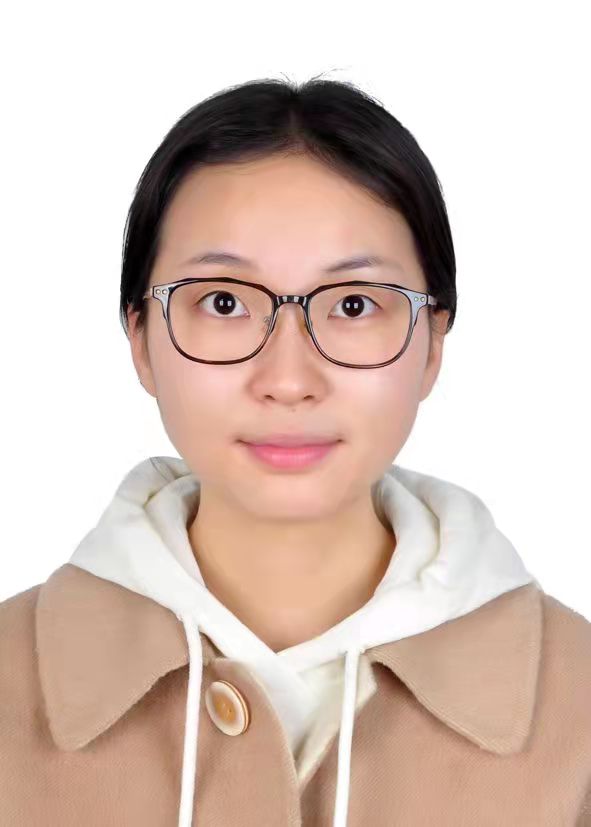}
\end{wrapfigure}
\textbf{Ruizhi Zhang} is currently pursuing the Ph.D. degree with the School of Mathematical Sciences, Beihang University, Beijing, China. Her current research interests include complex networks, link prediction and graph representation learning.

\subsection*{ }
\subsection*{ }
\subsection*{ }
\setlength\intextsep{0pt}
\begin{wrapfigure}{l}{25mm} 
	\includegraphics[width=1in,height=1.25in,clip,keepaspectratio]{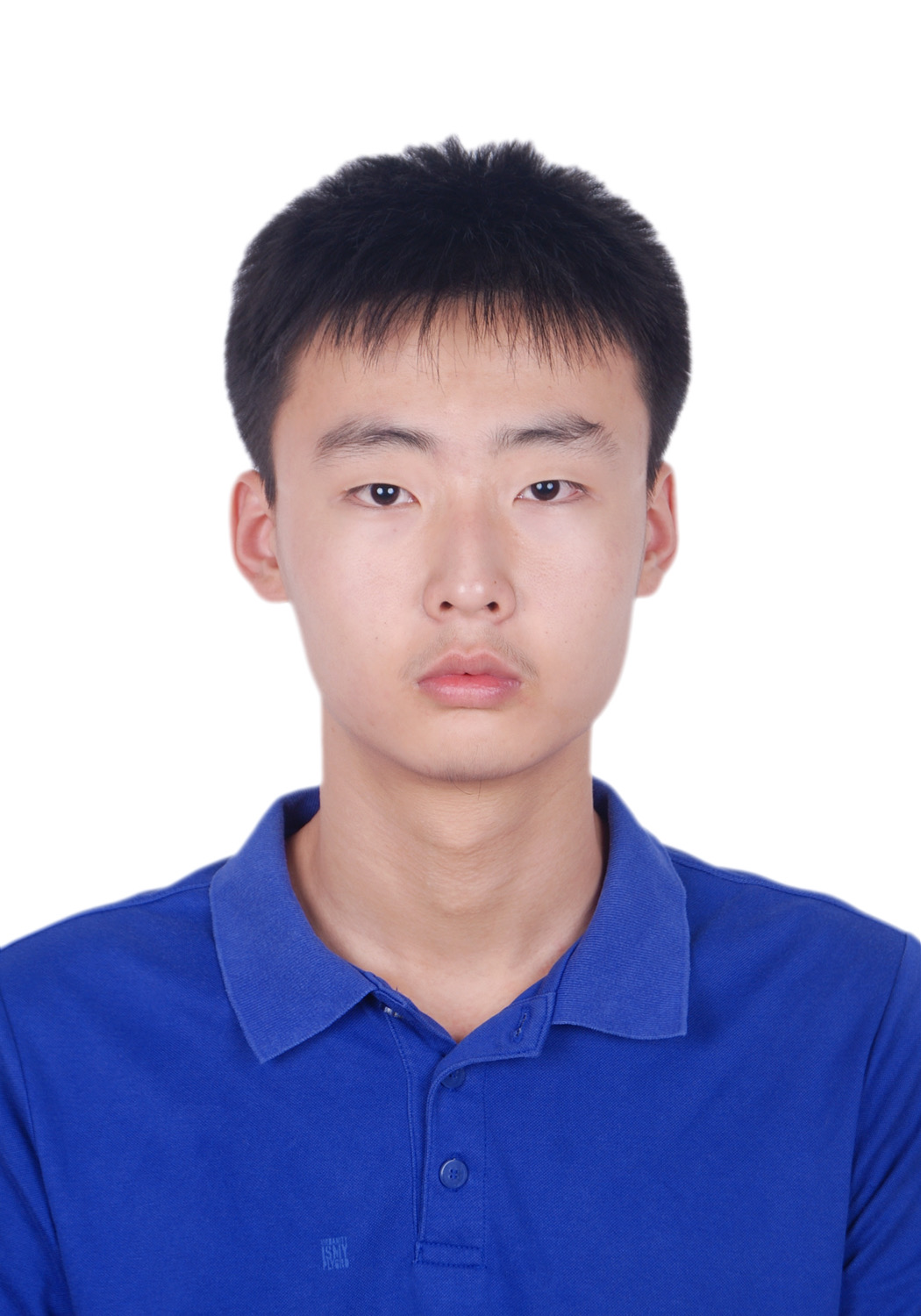}
\end{wrapfigure}
\textbf{Bowen Pang} received the B.S. degree from Beihang University, Beijing, China, in 2020. He is currently pursuing the Ph.D. degree with the School of Mathematical Sciences, Beihang University, Beijing, China. His current research interests include complex networks, neural networks, deep learning and time series analysis.

\subsection*{ }
\subsection*{ }
\setlength\intextsep{0pt}
\begin{wrapfigure}{l}{25mm} 
	\includegraphics[width=1in,height=1.25in,clip,keepaspectratio]{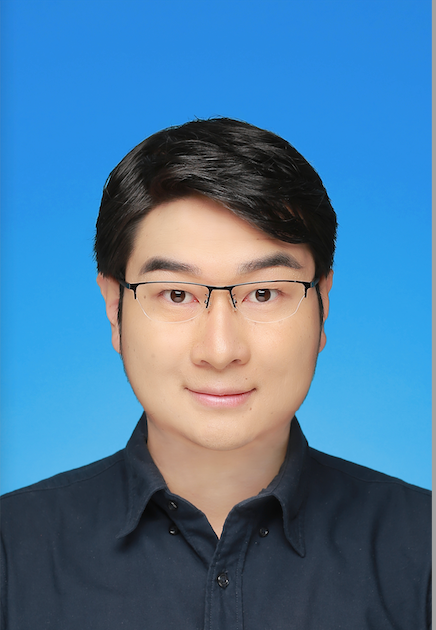}
\end{wrapfigure}
\textbf{Xiangnan Feng} received the Ph.D. degree in mathematics from the School of Mathematical Sciences, Beihang University,	Beijing, China, in 2020. He is currently a Postdoctoral Fellow with the Max Planck Institute for Human Development, Berlin, Germany. His research interests include complex networks, computing social science, and artificial intelligence.


\end{document}